\documentstyle[axodraw]{article}

\title{
\centerline{\normalsize hep-ph/9805291 \hfill SINP/TNP/98/11}
\bf Gravitational coupling of neutrinos in a medium}

\author{\bf Jos\'e F. Nieves\\
Department of Physics, P.O. Box 23343 \\
University of Puerto Rico, R\'{\i}o Piedras \\
Puerto Rico 00931-3343\\[12pt]
\bf Palash B. Pal\\ 
Saha Institute of Nuclear Physics, 1/AF Bidhan-Nagar \\
Calcutta 700064, India}
 
\date{May 1998}

\def\vec#1{\mbox{\boldmath $#1$}}

\begin{document}
\maketitle
\begin{abstract}

In a medium that contains electrons but not the other charged leptons,
such as normal matter, the gravitational interactions of neutrinos are
not the same for all the neutrino flavors. We calculate the leading
order matter-induced corrections to the neutrino gravitational interactions in
such a medium and consider some of their physical implications.

\end{abstract}

%
%
\section{Introduction}
\setcounter{equation}{0}
\label{sec:intro}
When neutrinos propagate through a medium, the effects of
the background particles can influence their properties in important ways.
The possible consequences of these effects have been the subject
of continuous research activity in recent years, largely motivated
by their implications in a variety of astrophysical and
cosmological contexts.  Some examples of the latter include
the original MSW mechanism \cite{msw} and its variations in the context
of the Solar Neutrino Problem \cite{solarnuprob}, and the explanation
of the large birth velocities of pulsars in terms of the 
asymmetric emission of neutrinos from the cooling protostar,
which is produced by the matter-enhanced
neutrino oscillations biased by the supernova's magnetic 
field~\cite{ks}.

Among the various approaches that exist to the study of the matter
effects on the propagation of neutrinos, the methods of Quantum
Statistical Field Theory (QSFT) have proven to be very useful ones.
These methods have been applied to reproduce the Wolfenstein formula
for the index of refraction of neutrinos in matter and to obtain
corrections to it \cite{nr,pp,jfn,dnt}.  In addition, they have been
used to determine the electromagnetic properties of a neutrino that
propagates in matter by means of the one-loop calculation of the
electromagnetic vertex function induced by the neutrino interactions
with the background particles \cite{dnp1}.  {}Furthermore, it was
observed in Ref.\ \cite{dnp1} that, in the presence of a static
magnetic field, the effective electromagnetic interactions of the
neutrinos produce an additional contribution to the neutrino index of
refraction which modifies the condition for resonant oscillations in
matter.  This effect is in fact the origin of the possible explanation
of the large birth velocities of pulsars mentioned above.

In all these situations, a common theme has been the
observation that the universality of the neutrino interactions
is broken due to the fact that normal matter contains electrons
but not the other charged leptons.  Therefore, in a medium
such as the Sun or a supernova, the electron neutrinos
on one hand, and the tau and muon neutrinos on the other,
are affected in different ways.  This fact implies
that in a such medium the GIM mechanism is not operative,
and it explains why the radiative decay of a neutrino
in a medium is greatly enhanced compared to the corresponding
rate in the vacuum~\cite{dnpradnudecay}.

The observation that the universality of the neutrino interactions is
broken by the background medium is a special case of a more general
concept. It is the notion that while the fundamental Lagrangian of the
theory is invariant under certain symmetry operations, a given
background medium may not be invariant under the same operations.
Thus, for example, normal matter is neither $CP$ nor $CPT$ asymmetric,
a fact that has interesting consequences for the propagation of
photons in a (chiral) medium such as a gas of neutrinos
\cite{mnp}. Taking this notion to the extreme, we can even think of
the medium as breaking Lorentz invariance since it specifies a
preferred frame of reference; i.e., that in which the medium is at
rest. In the context of QSFT, this breakdown of Lorentz invariance is
traded by an explicit dependence of the calculated physical properties
of the particles on the velocity four-vector of the background medium.
In this way, for example, the photon acquires a gauge invariant mass,
a left-handed neutrino acquires a chiral-invariant mass and a Majorana
neutrino acquires electromagnetic dipole moments.

The main observation of the present paper is that, in the
presence of medium, the breakdown of the universality
of the neutrino interactions includes  their gravitational
interactions as well.  In analogy with the fact that, in
a medium that contains electrons but no muons or taus,
the electron neutrinos have different electromagnetic
interactions than the muon or tau neutrinos,
their gravitational interactions also differ.
In this work we determine the effective gravitational
interactions of neutrinos in a matter background,
by calculating the one-loop contribution to the
neutrino stress-energy tensor, which is the gravitational analog
of the electromagnetic current.

We wish to clarify at the outset the following point.
The possibility that the observed deficit of solar
neutrinos can be ascribed to neutrino oscillations 
that are driven by flavor changing interactions
of gravitational origin has been proposed previously \cite{hl}.
In these works, it is postulated that the 
gravitational interaction of neutrinos has the same form
as the standard one, but with different coupling strengths
for each neutrino flavor.  This assumption violates
the equivalence principle in a fundamental way, and in fact
destroys the symmetry that makes it consistent to assume the existence
of a massless graviton in the vacuum.
On the other hand, no such fundamental breaking
of the equivalence principle is assumed in  the effect
that we describe in the present work.  Our calculations
are based on the standard model of particle interactions,
together with the commonly accepted linearized gravitational
coupling of fermions via the stress-energy tensor, with
a universal coupling strength.  The non-universal character
of the effective gravitational coupling of the neutrinos,
whose calculation is the aim of this work, is a consequence
of the flavor asymmetry of the medium and not of the fundamental
Lagrangian, as we have emphasized above.

We have found convenient to organize the presentation
in the following manner. In Section 2, we outline the
linearized theory of the gravitational coupling of fermions,
including the interaction terms with the $W$ and $Z$ gauge bosons.
This has been necessary because the formulas that are
commonly used and/or quoted in the literature 
are valid when the fermions that couple to the graviton
are on their mass-shell. Therefore, 
it is not appropriate to use them
for the internal fermion lines in the one-loop
diagrams that we need to consider.  Armed with the
preliminaries of  Section \ref{sec:gravcouplings}, 
the one-loop diagrams for the
induced gravitational interactions of the neutrinos in
a background composed of electrons and nucleons
are calculated in Section \ref{sec:indgrav}.
There we obtain the effective gravitational vertex function
as an integral over the momentum distribution functions
of the background particles.  We conclude that section
with some general remarks about the consistency
of the one-loop formulas, and in particular
we show explicitly that our result for the
effective  gravitational vertex of the
neutrinos, including the background-induced terms, is transverse
as it should be.
Using the one-loop formulas for the vertex function 
as a starting point, we  determine
in Section \ref{sec:refindex}
the modification to the neutrino index of refraction
in a medium in the presence of a static gravitational field. 
There we consider also some possible applications of these
results, giving some of the details in the Appendix,
and Section \ref{sec:conclusions} contains our conclusions.

%
%
\section{Tree-level gravitational couplings}
\setcounter{equation}{0}
\label{sec:gravcouplings}
It is well known that the linear, or weak field,  approximation
to the metric theory of gravity is equivalent to
a quantum particle description of gravitational interactions
in which the graviton emerges as a spin-2 quantum field 
coupled to the stress-energy tensor.
The formulas for the gravitational vertices is given and
discussed in many textbooks, at least for the common
cases of spin 0,1/2, and spin-1 (i.e., photons) particles.
However, the formulas that are customarily quoted are
given for the case in which the particles are on
their mass-shell, which is sufficient for the
applications that have been considered in the literature,
but not for the present one.
This is particularly true for the fermions.

In the one-loop calculation that we are considering, 
there are diagrams in which the graviton couples to internal electron
lines in a loop. For them, the on-shell form for the gravitational
vertex is not valid. Furthermore, there are other interaction
vertices that are unique to our case in hand. In this section
we consider all the  couplings that relevant to our
calculation in detail. We define the action $\cal A$ in presence of
gravitation by
	\begin{eqnarray}
{\cal A} = \int d^4x\; {\cal L} \,,
	\end{eqnarray}
and then look at the different terms in ${\cal L}$ which will be
relevant for us.

\subsection{Fermion couplings}
The Lagrangian for a free Dirac fermion of mass $m_f$ can be written
in the explicitly Hermitian form
\begin{eqnarray}\label{Lf0}
{\cal L}^{(f)}_0 = \left[
\frac{i}{2}\overline\psi \gamma^\mu \partial_\mu \psi 
+ {\rm h.c.} \right]
- m_f \overline\psi\psi  \,.
\end{eqnarray}
The Lagrangian in the presence of gravitational interactions
is obtained from this by making the replacement
$\gamma^\mu\partial_\mu\rightarrow\gamma^a {\cal D}_a$, where the
$\gamma^a$ denote the ordinary gamma matrices and
\begin{equation}\label{covder}
{\cal D}_a = v_a{}^\mu \left( \partial_\mu
-\frac{i}{4}\omega_{bc\mu}\sigma^{bc}\right)
\end{equation}
is the gravity-covariant derivative, with
$\sigma^{ab} =
\frac{i}{2}[\gamma^a,\gamma^b]$, and $v_a{}^\mu$ being the vierbein
vector fields. 
These are defined in a way that 
	\begin{eqnarray}\label{vierbeins}
\eta^{ab} v_a{}^\mu v_b^\nu =  g^{\mu\nu} \,, \qquad 
g_{\mu\nu} v_a{}^\mu v_b{}^\nu =  \eta_{ab} \,,
	\end{eqnarray}
where $g_{\mu\nu}$ is the space-time metric and 
$\eta_{ab}$ is the
flat space metric, which we take to be ${\rm diag}(1,-1,-1,-1)$.
The spin connection coefficients are given by
\begin{equation}\label{connection}
\omega_{ab\gamma} = v_{a\lambda}\left(
\partial_\gamma v_b{}^\lambda + 
\Gamma^\lambda{}_{\mu\gamma}v_b{}^\mu\right)\,,
\end{equation}
where $\Gamma_{\lambda\mu\gamma}$ are the Christoffel symbols.
This formula can be expressed explicitly in terms of the vierbein fields
in the form
\begin{eqnarray}\label{connection2}
\omega_{ab\gamma} & = & \frac{1}{2}
\left\{
v_a{}^\mu(\partial_\gamma v_{b\mu} -
\partial_\mu v_{b\gamma})  +
v_b{}^\mu(\partial_\mu v_{a\gamma} -
\partial_\gamma v_{a\mu})  +
v_a{}^\mu v_b{}^\nu v_{c\gamma}
(\partial_\nu v^c{}_\mu -
\partial_\mu v^c{}_\nu)
\right\}\,.
\end{eqnarray}
In addition to these changes, we have to include
the determinant of the matrix $v_{a\mu}$, which we denote
by ${\rm det}(v)$, as an overall factor.

Thus, in the absence of all interactions
except the gravitational ones, the Dirac Lagrangian is given by 
\begin{eqnarray}\label{Lfg}
{\cal L}^{(f)}_g = {\rm det}(v) \left\{ \left[
\frac{i}{2}\overline\psi\gamma^a v_a{}^\mu
\left( \partial_\mu
-\frac{i}{4}\omega_{bc\mu}\sigma^{bc}\right)
\psi  + {\rm h.c.} \right]
- m_f \overline\psi\psi\right\} \,.
\end{eqnarray}
With this construction, the term in the action 
corresponding to ${\cal L}^{(f)}_g$ 
is invariant under
general coordinate transformations, labeled by the greek
indices, and also under local Lorentz transformations
labeled by the latin indices.  

The linear theory of
the gravitational couplings is obtained by assuming
that, for weak gravitational fields, we can write
\begin{eqnarray}\label{glinear}
g_{\mu\nu} = \eta_{\mu\nu} + 2\kappa h_{\mu\nu}\,,
\end{eqnarray}
and treat the second term on the right side as a perturbation. The
quantity $\kappa$ is related to the Newton's constant $G$ through the
equation 
\begin{eqnarray}\label{fcoupling}
\kappa = \sqrt{8\pi G} \,,
\end{eqnarray}
in order that $h_{\mu\nu}$, identified with the graviton quantum field,
has the correctly normalized kinetic energy term in the Lagrangian.
The vierbeins cannot be determined uniquely from Eq.\ (\ref{vierbeins}).
They can be determined only
up to a local Lorentz transformation which, however, would leave
the action invariant.  Therefore, we can choose in particular
\begin{eqnarray}\label{vierbeinslinear}
v_{a\mu} = \eta_{a\mu} + \kappa h_{a\mu} \,,
\end{eqnarray}
which in turn gives
\begin{eqnarray}\label{detv}
{\rm det}(v) = 1 + \kappa \eta_{\mu\nu}h^{\mu\nu}\,.
\end{eqnarray}
Note that this also implies that
	\begin{eqnarray}
g^{\mu\nu} \simeq \eta^{\mu\nu} - 2\kappa h^{\mu\nu}
	\end{eqnarray}
since the matrix $g^{\mu\nu}$ is the inverse of the metric
$g_{\mu\nu}$ and from Eq.\ (\ref{vierbeins}), 
	\begin{eqnarray} 
v_a{}^\mu = \eta_a{}^\mu - \kappa h_a{}^\mu \,.
	\end{eqnarray}
We now substitute these relations
into Eq.\ (\ref{Lfg}), and keep only the terms that are at most
linear in $\kappa$.  Once this approximation is made, it is 
no longer necessary to distinguish between greek and
latin indices (since any difference between them would contribute only
to higher order in $\kappa$) and therefore we
write everything in terms of the greek
indices from now on. {}From Eq.\ (\ref{connection2}) we obtain
\begin{equation}\label{connectionlinear}
\omega_{\mu\nu\gamma} = \frac{1}{2}(\partial_\nu h_{\mu\gamma}
- \partial_\mu h_{\gamma\nu})\,,
\end{equation}
and it then follows that the
term involving the $\sigma$-matrices in Eq.\ (\ref{Lfg})
drops out because it is proportional to the quantity
\begin{equation}\label{Cabc}
\left\{\gamma_\mu,\sigma_{\lambda\nu}\right\} 
+ (\mu\leftrightarrow \nu)  = 0 \,.
\end{equation}
{}From the other terms we then
obtain the gravitational interaction Lagrangian
of the fermion in the form
\begin{eqnarray}\label{Lffh}
{\cal L}^{(ff)}_{h} = -\kappa h^{\mu\nu} (x) \widehat T^{(f)}_{\mu\nu} (x) \,,
\end{eqnarray}
where the stress-energy 
tensor operator $\widehat T_{\mu\nu}^{(f)}$ for the fermion field is given
by
\begin{eqnarray}\label{stresstensor}
\widehat T^{(f)}_{\mu\nu} (x) = \left\{
{i\over 4} \overline \psi(x) \left[\gamma_\mu \partial_\nu + \gamma_\nu
\partial_\mu \right] \psi(x) + {\rm h.c.} \right\} -
\eta_{\mu\nu} {\cal L}_0^{(f)} (x) \,.
\end{eqnarray}
{}From Eqs.\ (\ref{Lffh}) and (\ref{stresstensor}) it follows that the term
corresponding to the gravitational
fermion vertex in a Feynman diagram is
$-i\kappa V_{\mu\nu}^{(f)}$, where
\begin{eqnarray}\label{Vmunu}
V_{\mu\nu}^{(f)} (p,p') = \frac{1}{4} \left[
\gamma_\mu(p + p')_\nu + 
\gamma_\nu(p + p')_\mu \right]
- \frac{1}{2}\eta_{\mu\nu}
\left[(\rlap/ p - m_f) + (\rlap/ p' - m_f) \right] \,.
\end{eqnarray}

In the above considerations we have assumed that the fermion
$f$ is a Dirac particle with a given mass.  
On the other hand, it is easy to infer by inspection that
similar arguments yield the formula
\begin{eqnarray} \label{Vneutrino}
V_{\mu\nu}^{(\nu)} (k,k') = \frac{1}{4} \left[
\gamma_\mu(k + k')_\nu + 
\gamma_\nu(k + k')_\mu \right] L
- \frac{1}{2}\eta_{\mu\nu}
\left[ \rlap/ k + \rlap/k \,' \right] L 
\end{eqnarray}
for the case of a chiral, left-handed neutrino.

Notice that if we take the matrix
element of the operator $\widehat T_{\mu\nu}^{(f)}(0)$ 
between on-shell fermion states, with incoming 
and outgoing momenta $p$ and $p'$ respectively,
the term proportional to $\eta_{\mu\nu}$ in 
Eq.\ (\ref{stresstensor}) gives no contribution and we obtain
\begin{eqnarray}\label{Tonshell}
\left< f(p')\left| \widehat T^{(f)}_{\mu\nu} (0) \right|f(p)\right> =
\frac{1}{4}\overline u(p')\left\{
\gamma_\mu(p + p')_\nu + 
\gamma_\nu(p + p')_\mu\right\}u(p)\,,
\end{eqnarray}
which is the expression that is quoted in textbooks~\cite{scadron}. 
However, for the purposes of the one-loop calculation that we carry
out in Section \ref{sec:indgrav}, we need to use the vertex for
off-shell fermions given in Eq.\ (\ref{Vmunu}).  In particular, as
we will show in Section \ref{sec:indgrav},  the one-loop
calculation of the effective gravitational
vertex of the neutrino gives a result 
that satisfies the transversality condition
provided that  the 
term proportional to $\eta_{\mu\nu}$ in Eq.\ (\ref{Vmunu})
is included in the calculation of the loop diagrams.

\subsection{Fermion and $W$ boson couplings}
The interactions that drive the effective gravitational
vertex of the neutrinos, are the standard weak
interactions with the particles of the background.
Let us consider the charged-current interactions first.
In the presence of a gravitational field,
the interaction term $\overline e_L\gamma^\mu W_\mu\nu_L$
in the Lagrangian is modified according to
	\begin{eqnarray}\label{Lchg}
{\cal L}^{\rm (cc)}_g = {\rm det}(v)\left\{ -\frac{g}{\sqrt{2}}\overline
e_L \gamma^a \nu_L v_{a\mu} W^\mu + {\rm h.c.} \right\} \,.
	\end{eqnarray}
In the linear approximation given by Eqs.\ (\ref{vierbeinslinear})
and (\ref{detv}), this becomes
	\begin{eqnarray}\label{Lchg2}
{\cal L}^{\rm (cc)}_g = {\cal L}^{\rm (cc)}_0 + {\cal L}^{\rm (cc)}_h \,,
	\end{eqnarray}
where ${\cal L}^{\rm (cc)}_0$ is the standard charged-current interaction
Lagrangian
\begin{eqnarray}\label{Lch}
{\cal L}^{\rm (cc)}_0 = -\frac{g}{\sqrt{2}} W^\mu \overline
e_L \gamma_\mu \nu_L + {\rm h.c.} \,, 
\end{eqnarray}
while
\begin{eqnarray}\label{LWenuh}
{\cal L}^{\rm (cc)}_h &=& -\kappa \frac{g}{\sqrt{2}} h^{\mu\nu} \left[
\eta_{\mu\nu} W_\alpha \overline e_L \gamma^\alpha\nu_L +
\frac{1}{2}\left(W_\mu\overline e_L\gamma_\nu\nu_L + W_\nu\overline e_L
\gamma_\mu\nu_L\right) + {\rm h.c.} \right] \nonumber\\ 
&=& -\kappa \frac{g}{\sqrt{2}} \;a_{\mu\nu\lambda\rho} \; \overline
e_L \gamma^\lambda \nu_L \, W^\rho h^{\mu\nu} + {\rm h.c.} \,,
\end{eqnarray}
where
	\begin{eqnarray}\label{defa}
a_{\mu\nu\lambda\rho} = \eta_{\mu\nu}\eta_{\lambda\rho} + {1\over 2}
\left[ 
\eta_{\mu\lambda}\eta_{\nu\rho} +
\eta_{\mu\rho}\eta_{\nu\lambda} 
\right] \,.
	\end{eqnarray}

\subsection{$W$ boson couplings}
The gravitational vertex for photons is well known. {}For
the $W$ boson, the only difference comes from the mass
term in the kinetic energy part of the Lagrangian.
Indeed, from the usual expression for the
kinetic energy of the $W$ vector boson,
\begin{eqnarray}\label{LWg}
{\cal L}^{(W)}_g = {\rm det}(v) \left\{ -\frac{1}{2} W^{\ast}_{\mu\nu}
W^{\mu\nu} - M_W^2 W^\ast_\mu W^\mu\right\} \,,
\end{eqnarray}
and then making the substitutions
given in Eqs.\ (\ref{glinear}) and (\ref{detv}), we arrive at 
\begin{eqnarray}\label{LWg2}
{\cal L}^{(W)}_g = {\cal L}^{(W)} + {\cal L}^{(WW)}_h \,,
\end{eqnarray}
where ${\cal L}^{(W)}$ is the standard form of the 
kinetic energy for the free $W$ boson, while
\begin{eqnarray}\label{LWWh}
{\cal L}^{(WW)}_h = -\kappa h^{\mu\nu} \left[ \left( 
\frac{1}{2} W^\ast_{\alpha\beta} 
W^{\alpha\beta} - M_W^2 W^\ast_\alpha W^\alpha\right)\eta_{\mu\nu} +
\left(W^\ast_{\mu\alpha}W^\alpha_\nu - 
M_W^2 W^\ast_\mu W_\nu + {\rm h.c.}\right) \right]
\end{eqnarray}
gives the gravitational vertex.

\begin{figure}
\begin{center}
%
%
\begin{picture}(180,130)(-90,-65)
\Text(35,-35)[ct]{\large\bf (A)}
\ArrowLine(80,0)(40,0)
\Text(60,-10)[c]{$\nu_e(k)$}
\ArrowLine(40,0)(0,0)
\Text(20,-10)[c]{$e(p)$}
\ArrowLine(0,0)(-40,0)
\Text(-20,-10)[c]{$e(p-q)$}
\ArrowLine(-40,0)(-80,0)
\Text(-60,-10)[c]{$\nu_e(k')$}
\Photon(0,0)(0,-45){2}{4}
\Photon(0,0)(0,-45){-2}{4}
\Text(0,-50)[l]{$q$}
\PhotonArc(0,0)(40,0,180){4}{7.5}
\Text(0,50)[c]{$W$}
\end{picture}
%
%
\begin{picture}(180,130)(-90,-65)
\Text(0,-35)[ct]{\large\bf (B)}
\ArrowLine(80,0)(40,0)
\Text(60,-10)[c]{$\nu_e(k)$}
\ArrowLine(40,0)(-40,0)
\Text(0,-10)[c]{$e(p)$}
\ArrowLine(-40,0)(-80,0)
\Text(-60,-10)[cr]{$\nu_e(k')$}
\Photon(0,44)(0,80){2}{4}
\Photon(0,44)(0,80){-2}{4}
\Text(0,85)[bl]{$q$}
\PhotonArc(0,0)(40,0,180){4}{6.5}
\Text(40,40)[c]{$W$}
\Text(-40,40)[c]{$W$}
\end{picture}
%
%
\begin{picture}(180,130)(-90,-65)
\Text(0,-35)[ct]{\large\bf (C)}
\ArrowLine(80,0)(40,0)
\Text(60,-10)[c]{$\nu_e(k)$}
\ArrowLine(40,0)(-40,0)
\Text(0,-10)[c]{$e(p)$}
\ArrowLine(-40,0)(-80,0)
\Text(-60,-10)[c]{$\nu_e(k')$}
\PhotonArc(0,0)(40,0,180){4}{7.5}
\Text(0,50)[c]{$W$}
\Photon(40,0)(40,-50){2}{4}
\Photon(40,0)(40,-50){-2}{4}
\Text(40,-55)[l]{$q$}
\end{picture}
%
%
%
\begin{picture}(180,130)(-90,-65)
\Text(0,-35)[ct]{\large\bf (D)}
\ArrowLine(80,0)(40,0)
\Text(60,-10)[c]{$\nu_e(k)$}
\ArrowLine(40,0)(-40,0)
\Text(0,-10)[c]{$e(p)$}
\ArrowLine(-40,0)(-80,0)
\Text(-60,-10)[c]{$\nu_e(k')$}
\PhotonArc(0,0)(40,0,180){4}{7.5}
\Text(0,50)[c]{$W$}
\Photon(-40,0)(-40,-50){2}{4}
\Photon(-40,0)(-40,-50){-2}{4}
\Text(-40,-55)[r]{$q$}
\end{picture}

\caption[]{\sf $W$ exchange diagrams for the one-loop contribution
to the $\nu_e$ gravitational vertex in a background of electrons. The
braided line represents the graviton.
\label{fig:wdiagrams}}

\end{center}
\end{figure}
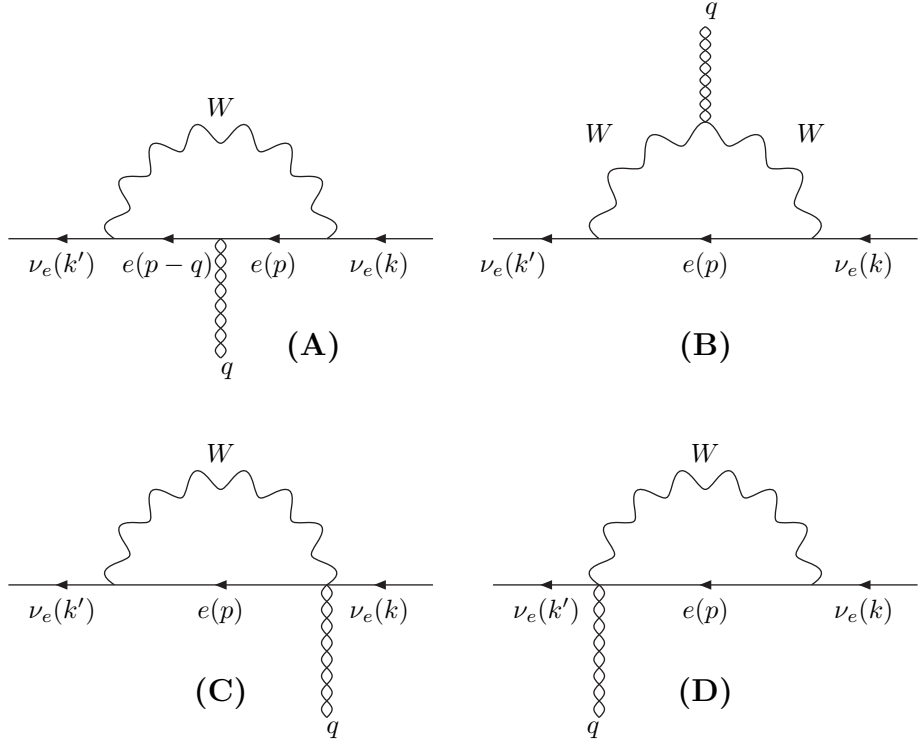

Of particular importance for us are the terms proportional
to the $W$ mass. The reason is the following.  In the analogous
calculation of the charged-current contribution to the
electromagnetic neutrino vertex in a medium, the dominant 
term is of order $1/M_W^2$ and it arises from the diagram
that corresponds to diagram (A) in Fig.~\ref{fig:wdiagrams},
in which the photon couples to the electron line in the loop.
The diagram in which the photon couples to the internal $W$ line,
corresponding to diagram (B), 
is of order $1/M_W^4$, and therefore it is negligible.
In the present case, because of the presence of the term 
that is proportional to the $W$ mass in the coupling of the $W$
to the graviton,
diagram (B) gives a contribution also of order $1/M_W^2$ that
must be taken into account. As far as these terms only are concerned,
we can then replace Eq.\ (\ref{LWWh}) by
	\begin{eqnarray}
{\cal L}^{(WW)}_h = \kappa M_W^2 a'_{\mu\nu\lambda\rho} \;  h^{\mu\nu}
W^\lambda W^{*\rho} \,,
	\end{eqnarray}
where
	\begin{eqnarray}\label{defa'}
a'_{\mu\nu\lambda\rho} = 
\eta_{\mu\nu} \eta_{\lambda\rho} + \eta_{\mu\lambda} \eta_{\nu\rho} + 
\eta_{\mu\rho} \eta_{\nu\lambda} \,.
	\end{eqnarray}

\subsection{Including the neutral-current couplings}
The charged current interaction discussed above are relevant only for
the $\nu_e$'s. The $\nu_\mu$'s and the $\nu_\tau$'s will interact with
the electrons only through the neutral current. Moreover, a normal
background contains nucleons as well, which interact with all
neutrinos via the neutral current. To take these interactions into
account, we consider the neutral current couplings of the
$Z$-boson. In absence of gravitation, these are
	\begin{eqnarray}\label{LZint}
{\cal L}^{\rm (nc)}_0 = -g_ZZ^\mu\left[\sum_{i = e,\mu,\tau}
\overline\nu_{iL}\gamma_\mu \nu_{iL} +
\sum_{f = e,p,n}
\overline f\gamma_\mu(X_f + Y_f\gamma_5)f\right] \,,
	\end{eqnarray}
where, in the standard model,
\begin{eqnarray}\label{Zcouplings}
g_Z & = & g/(2\cos\theta_W) \nonumber\\
-X_e = X_p  & = & \frac{1}{2} - 2\sin^2\theta_W \nonumber\\
X_n & = & -\frac{1}{2} \nonumber\\
Y_e & = & \frac{1}{2} \nonumber\\
Y_n = -Y_p & = & \frac{1}{2}g_A \,,
\end{eqnarray}
with $g_A = 1.26$ being the renormalization constant
of the axial-vector current of the nucleon.
It is also useful to remember that, in the standard model,
\begin{equation}\label{GFdef}
\frac{g^2_Z}{M^2_Z} = \frac{g^2}{4M_W^2} = \sqrt{2}G_F \,.
\end{equation}
Following the arguments that were used for the charged current, we
find that in presence of gravitation, these interactions will be
modified to
	\begin{eqnarray}\label{Lncg2}
{\cal L}^{\rm (nc)}_g = {\cal L}^{\rm (nc)}_0 + {\cal L}^{\rm (nc)}_h \,,
	\end{eqnarray}
where ${\cal L}^{\rm (nc)}_0$ has been given above in Eq.\ (\ref{LZint}),
and 
	\begin{eqnarray}\label{Lnch}
{\cal L}^{\rm (nc)}_h = -\kappa g_Z \;a_{\mu\nu\lambda\rho} \; 
\left[\sum_{i = e,\mu,\tau}
\overline\nu_{iL}\gamma^\lambda \nu_{iL} +
\sum_{f = e,p,n}
\overline f\gamma^\lambda (X_f + Y_f\gamma_5)f\right] 
Z^\rho h^{\mu\nu} \,,
	\end{eqnarray}
where $a_{\mu\nu\lambda\rho}$ has been defined in Eq.\
(\ref{defa}). Also, there are terms which comes from the Lagrangian of
the pure $Z$-boson which give interactions with the graviton and are
proportional to $M_Z^2$. These are
	\begin{eqnarray}\label{LZZh}
{\cal L}^{(ZZ)}_h = {1\over 2}\kappa M_Z^2 a'_{\mu\nu\lambda\rho} \;
h^{\mu\nu} Z^\lambda Z^\rho \,,
	\end{eqnarray}
where $a'_{\mu\nu\lambda\rho}$ is the tensor defined in Eq.\
(\ref{defa'}). 

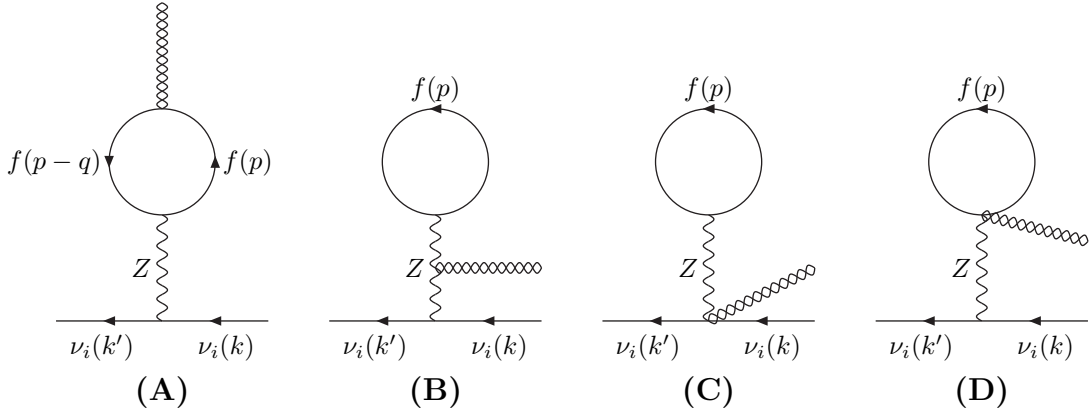
\begin{figure}
\begin{center}
%
%
\begin{picture}(100,170)(-50,-30)
\Text(0,-30)[cb]{\large\bf (A)}
\ArrowLine(40,0)(0,0)
\Text(35,-10)[cr]{$\nu_i(k)$}
\ArrowLine(0,0)(-40,0)
\Text(-35,-10)[cl]{$\nu_i(k^\prime)$}
\Photon(0,0)(0,40){2}{6}
\Text(-4,20)[r]{$Z$}
\ArrowArc(0,60)(20,90,270)
\ArrowArc(0,60)(20,-90,90)
\Text(23,60)[l]{$f(p)$}
\Text(-23,60)[r]{$f(p - q)$}
\Photon(0,80)(0,120){2}{6}
\Photon(0,80)(0,120){-2}{6}
\end{picture}
%
%
\begin{picture}(100,170)(-50,-30)
\Text(0,-30)[cb]{\large\bf (B)}
\ArrowLine(40,0)(0,0)
\Text(35,-10)[cr]{$\nu_i(k)$}
\ArrowLine(0,0)(-40,0)
\Text(-35,-10)[cl]{$\nu_i(k^\prime)$}
\Photon(0,0)(0,40){2}{6}
\Text(-4,20)[r]{$Z$}
\ArrowArc(0,60)(20,-90,270)
\Text(0,85)[b]{$f(p)$}
\Photon(0,20)(40,20){2}{6}
\Photon(0,20)(40,20){-2}{6}
\end{picture}
%
%
\begin{picture}(100,170)(-50,-30)
\Text(0,-30)[cb]{\large\bf (C)}
\ArrowLine(40,0)(0,0)
\Text(35,-10)[cr]{$\nu_i(k)$}
\ArrowLine(0,0)(-40,0)
\Text(-35,-10)[cl]{$\nu_i(k^\prime)$}
\Photon(0,0)(0,40){2}{6}
\Text(-4,20)[r]{$Z$}
\ArrowArc(0,60)(20,-90,270)
\Text(0,85)[b]{$f(p)$}
\Photon(0,0)(40,20){2}{6}
\Photon(0,0)(40,20){-2}{6}
\end{picture}
%
%
\begin{picture}(100,170)(-50,-30)
\Text(0,-30)[cb]{\large\bf (D)}
\ArrowLine(40,0)(0,0)
\Text(35,-10)[cr]{$\nu_i(k)$}
\ArrowLine(0,0)(-40,0)
\Text(-35,-10)[cl]{$\nu_i(k^\prime)$}
\Photon(0,0)(0,40){2}{6}
\Text(-4,20)[r]{$Z$}
\ArrowArc(0,60)(20,-90,270)
\Text(0,85)[b]{$f(p)$}
\Photon(0,40)(40,30){2}{6}
\Photon(0,40)(40,30){-2}{6}
\end{picture}

\end{center}

\caption[]{\sf $Z$-exchange diagrams for the one-loop contribution
to the gravitational vertex of any neutrino flavor
($i = e,\mu,\tau$) in a background of electrons and
nucleons. 
\label{fig:zdiagrams}}

\end{figure}

%
%
\section{Induced gravitational vertex of neutrinos}
\setcounter{equation}{0}
\label{sec:indgrav}
We now consider the background contributions
to the neutrino gravitational vertex.
We denote the proper vertex function for off-shell neutrinos,
including the background-induced contributions, 
by $\Gamma^{(\nu)}_{\mu\nu}(k,k^\prime)$.  It is
is defined such that the matrix element of 
the total stress-energy tensor operator
$\widehat T_{\mu\nu}(x)$,
between incoming and outgoing neutrino states
with momenta $k$ and $k^\prime$ respectively, is given
by
\begin{equation}\label{defGameff}
\left< \nu(k')\left| \widehat T_{\mu\nu} (0) \right|\nu(k)\right> =
\overline u_L(k^\prime)\Gamma^{(\nu)}_{\mu\nu}(k,k^\prime)u_L(k)\,.
\end{equation}
It is useful to divide the relevant one-loop diagrams
into the two sets given in
Figs.\ \ref{fig:wdiagrams} and \ref{fig:zdiagrams},
according  to whether they involve the $W$ or the $Z$ boson.
In the diagrams, and in the formulas that follow,
$q$ stands for the outgoing graviton momentum
\begin{eqnarray}\label{q}
q = k - k' \,.
\end{eqnarray}
The additional diagrams in which the graviton line comes out from
one of the external neutrino legs are not shown,
since they are 1-particle reducible and do not
contribute to $\Gamma^{(\nu)}_{\mu\nu}$.  
The proper way to take them into account in the calculation 
of the amplitude for any given process, is by
choosing the external neutrino spinor $u_L(k)$ to be 
the (properly normalized) solution
of the effective Dirac equation for the
propagating neutrino mode in the medium, instead of the spinor
representing the free-particle solution of the equation in the vacuum.
This will be discussed in more detail in Section\ \ref{subsec:transv}.

As commented earlier, the $W$-exchange diagrams of Fig.\
\ref{fig:wdiagrams} contribute only to the gravitational coupling of
$\nu_e$. In contrast, the diagrams shown in Fig.\
\ref{fig:zdiagrams} contribute equally to the gravitational vertex of
all the neutrino flavors $\nu_{e,\mu,\tau}$, and therefore are not
relevant in phenomena that involve transitions between the
standard, weak SU(2)-doublet neutrinos. However,
in processes in which the so-called sterile neutrinos
participate, these flavor-diagonal contributions are important.
Moreover, we consider them first since the results for the $W$-diagrams
can be easily obtained from the corresponding ones for 
the $Z$-diagrams by making some simple substitutions.

\subsection{The $Z$-mediated diagrams}
\subsubsection{Diagram (A)}
This contribution, which will be marked by the superscript 2A, is
given by
\begin{eqnarray}
-i\kappa \Gamma_{\mu\nu}^{(2A)} = -(-ig_Z)^2(-i\kappa)
\frac{i}{M_Z^2}\gamma^\lambda L \sum_{f=e,p,n} 
\int {d^4p \over (2\pi)^4} \; {\rm Tr} \left[
\gamma_\lambda (X_f + Y_f\gamma_5)\, iS_F^{(f)} (p')
V_{\mu\nu}^{(f)} (p,p') iS_F^{(f)} (p)
\right] \,,
	\end{eqnarray}
or equivalently
	\begin{eqnarray}\label{Gamma2a}
\Gamma_{\mu\nu}^{(2A)} = 
{ig_Z^2\over M_Z^2} \, \gamma^\lambda L \sum_{f=e,p,n} 
\int {d^4p \over (2\pi)^4} \; {\rm Tr} \left[
\gamma_\lambda (X_f + Y_f\gamma_5)\, iS_F^{(f)} (p') 
V_{\mu\nu}^{(f)} (p,p') \, iS_F^{(f)} (p)
\right] \,,
	\end{eqnarray}
In these expressions we have introduced the shorthand
\begin{eqnarray}\label{pprime}
p' \equiv p - q \,,
\end{eqnarray}
which will be used throughout the calculations, and
we have neglected the momentum dependence of the $Z$ 
propagator in anticipation
of the fact that we are interested only in the terms that are linear
in the Fermi constant $G_F$.  
Further, $S_F^{(f)}$ is the thermal fermion propagator which is given by
\begin{eqnarray}\label{SFf}
iS_F^{(f)}(p) = (\rlap/{p} + m_f)\left[
\frac{i}{p^2 - m_f^2 + i\epsilon} - 2\pi\delta(p^2 -
m_f^2)\eta_f(p)\right] \,, 
\end{eqnarray}
where
\begin{eqnarray}\label{etaf}
\eta_f(p) = \frac{\theta(p\cdot v)}{e^{\beta(p\cdot v - \mu_f)} + 1}
+ \frac{\theta(-p\cdot v)}{e^{-\beta(p\cdot v + \mu_f)} + 1}
\end{eqnarray}
with $\beta$ being the inverse temperature, $\mu_f$ the
chemical potential and $v^\mu$ the velocity
four-vector of the medium \cite{footnote1}.
In the frame in which the medium 
is at rest, $v^\mu$ has components 
\begin{eqnarray}\label{vatrest}
v^\mu = (1,\vec 0) \,.
\end{eqnarray}
When the expressions for $S_F^{(f)}(p)$ and $S_F^{(f)}(p')$
given Eq.\ (\ref{SFf}) are substituted in Eq.\ (\ref{Gamma2a}), three 
kinds of term are generated, according to whether they
contain none, one or two factors of $\eta_f$.
We are interested in the terms with at least one
factor of $\eta_f$, since that is where the background
dependence of the neutrino gravitational vertex will come from.
However, the terms with two factors of $\eta_f$
are relevant only in the calculation of the absorptive part of the amplitude,
which we are not considering here, and
therefore the important terms for us are the ones having a single factor of
$\eta_f$. Denoting their contribution to the vertex
function by $\Gamma'{}_{\mu\nu}^{(2A)}$
we then obtain
	\begin{eqnarray}\label{Gam2A}
\Gamma'{}_{\mu\nu}^{(2A)} 
&=& {g_Z^2 \over M_Z^2}\; \gamma^\lambda L 
\sum_{f=e,p,n} 
\int {d^4p \over (2\pi)^3} \; {\rm Tr} \left[
\gamma_\lambda (X_f + Y_f\gamma_5) 
A_{\mu\nu} \right] \nonumber\\* 
&& \qquad \qquad \times \left( {\delta(p^2-m_f^2) \eta_f(p)
\over p'^2-m_f^2} + {\delta(p'^2-m_f^2) \eta_f(p')
\over p^2-m_f^2} \right) \,,
	\end{eqnarray}
where
	\begin{eqnarray}\label{Amunu}
A_{\mu\nu}^{(f)} = (\rlap/p'+m_f) V_{\mu\nu}^{(f)} (p,p') (\rlap/p+m_f) \,.
	\end{eqnarray}
Using for $V_{\mu\nu}^{(f)}$ the expression given
in Eq.\ (\ref{Vmunu}), we have
\begin{eqnarray}\label{Amunu2}
A_{\mu\nu}^{(f)} = a_{\mu\nu}^{(f)} -\frac{1}{2}\eta_{\mu\nu}
\left[(\rlap/p \,' + m_f)(p^2 - m_f^2) + 
(\rlap/p + m_f) (p'{}^2 - m_f^2)\right] 
\end{eqnarray}
where
\begin{eqnarray}\label{amunu}
a_{\mu\nu}^{(f)} = (\rlap/p\,' + m_f)
\frac{1}{4}\left\{\gamma_\mu(p + p')_\nu + 
\gamma_\nu(p + p')_\mu\right\}
(\rlap/p + m_f) \,.
\end{eqnarray}
When Eq.\ (\ref{Amunu2}) is substituted in Eq.\ (\ref{Gam2A}), 
the terms that involve the factor of $\eta_{\mu\nu}$
are reduced very simply, and the result can be expressed
in the form
\begin{eqnarray}\label{Gam2Afinal}
\Gamma'{}^{(2A)}_{\mu\nu} = \Lambda^{(Z)}_{\mu\nu} -
b^{(Z)} \rlap/{v} L\eta_{\mu\nu} \,,
\end{eqnarray}
where $\Lambda^{(Z)}_{\mu\nu}$ is given by the same
expression given in Eq.\ (\ref{Gam2A}) but with the replacement
$A_{\mu\nu}\rightarrow a_{\mu\nu}$, and 
\begin{equation}\label{bZ}
b^{(Z)} = \sum_{f} X_fb_f \,,
\end{equation}
with
\begin{equation}\label{b}
b_f = 4\sqrt{2}G_F
\int\frac{d^4p}{(2\pi)^3}\delta(p^2- m_f^2) \eta_f(p) \; p\cdot v \,.
\end{equation}
In arriving at Eq.\ (\ref{Gam2Afinal}) we have used the formula
\begin{equation}\label{intp}
4\sqrt{2}G_F\int\frac{d^4p}{(2\pi)^3}
\delta(p^2 - m_f^2)\eta_f(p)p_\mu = b_fv_\mu \,,
\end{equation}
and the relation $\sqrt{2}G_F = g_Z^2/M_Z^2$.

It is convenient to introduce
the particle and antiparticle momentum distribution functions
\begin{eqnarray}\label{fe}
f_{f,\overline f}(p_f) = \frac{1}{e^{\beta(p_f\cdot v \mp \mu_f)}}\,, 
\end{eqnarray}
where the upper and the lower signs hold for the particle and the
antiparticle respectively, and
\begin{eqnarray}\label{pErel}
p^\mu_f = (E_f,\vec P)\,, \qquad E_f = \sqrt{\vec P^2 + m_f^2} \,.
\end{eqnarray}
The corresponding total number densities
are given by
\begin{equation}\label{ne}
n_f = 2 \int {d^3P \over (2\pi)^3} \; f_f \,, \qquad 
n_{\bar f} = 2 \int {d^3P \over (2\pi)^3} \; f_{\bar f} \,,
\end{equation}
and in terms of them
Eq.\ (\ref{b}) can then be written in the form
\begin{eqnarray}\label{b2}
b_f = \sqrt{2}G_F(n_f + n_{\overline f}) \,.
\end{eqnarray}

The expression for
$\Lambda^{(Z)}_{\mu\nu}$  can be simplified by taking the traces
and making change of integration
variable $p \rightarrow p+q$ in the term containing
the factor of $\eta_f(p')$.  In this way we obtain
\begin{eqnarray}\label{LambdaZ}
\Lambda^{(Z)}_{\mu\nu} &=& 
\frac{g_Z^2}{M_Z^2}(\gamma^\lambda L) \sum_f \int\frac{d^4p}{(2\pi)^3}
\delta(p^2 - m_e^2)\eta_f(p) \nonumber\\*
&& \quad \quad \times \left\{
\frac{X_f N^{(1)}_{\mu\nu\lambda}(p,q) -
Y_f N^{(2)}_{\mu\nu\lambda}(p,q)}{q^2 - 2p\cdot q} +
\frac{X_f N^{(1)}_{\mu\nu\lambda}(p,-q) + 
Y_f N^{(2)}_{\mu\nu\lambda}(p,-q)}{q^2 + 2p\cdot q}
\right\}
\end{eqnarray}
where
\begin{eqnarray}\label{N}
N^{(1)}_{\mu\nu\lambda}(p,q) & \equiv &
(2p - q)_\mu\left[2p_\nu p_\lambda
- (p_\lambda q_\nu + q_\lambda p_\nu) + (p\cdot q)\eta_{\lambda\nu}\right]
+ (\mu\leftrightarrow\nu) \nonumber\\
N^{(2)}_{\mu\nu\lambda}(p,q) & \equiv &
(2p - q)_\mu i\epsilon_{\nu\lambda\rho\sigma}q^\rho p^\sigma +
(\mu\leftrightarrow\nu) \,.
\end{eqnarray}
It is useful to observe that these quantities satisfy the relations
\begin{eqnarray}\label{Srelations}
N^{(1,2)}_{\mu\nu\lambda}(-p,-q) = - N^{(1,2)}_{\mu\nu\lambda}(p,q) \,,
\end{eqnarray}
from which other similar relations can be obtained.
Using them, after performing the $p^0$ integration
in Eq.\ (\ref{LambdaZ}) we obtain finally
	\begin{eqnarray}\label{LambdaZfinal}
\Lambda^{(Z)}_{\mu\nu} = 
\frac{g_Z^2}{M_Z^2}(\gamma^\lambda L)
\sum_{f}\int\frac{d^3P}{2E_f(2\pi)^3}
\left\{X_f(f_f - f_{\overline f})\left[
\frac{N^{(1)}_{\mu\nu\lambda}(p_f,q)}{q^2 - 2p_f\cdot q} + 
(q\rightarrow -q)\right]\right.\nonumber\\
- 
\left. Y_f(f_f + f_{\overline f})\left[
\frac{N^{(2)}_{\mu\nu\lambda}(p_f,q)}{q^2 - 2p_f\cdot q} -
(q\rightarrow -q)\right]
\right\} \,.
	\end{eqnarray}
Further reduction of Eq.\ (\ref{LambdaZfinal}) is not
possible without making some assumption about the
conditions of the electron gas, in order to be able
to carry out the required integrations.  Before turning
to that, we consider the contributions from the other
diagrams, that must be added to Eq.\ (\ref{Gam2Afinal}) to
yield the full one-loop result for the induced vertex function.

\subsubsection{Diagram (B)}
Because this diagram contains two $Z$-boson propagators,
we retain only the terms proportional to $M_Z^2$ 
from the $ZZh$ vertex given in Eq.\ (\ref{LZZh}), since the rest
will yield results proportional to $1/M_Z^4$.  For the same
reason, we also neglect the momentum dependence of
the two $Z$ propagators in the diagram.  In this way
we then obtain
\begin{eqnarray} 
-i\kappa\Gamma^{(2B)}_{\mu\nu} 
= - (-ig_Z)^2
\left(\frac{i}{M_Z^2}\right)^2
(i\kappa M_Z^2 a'_{\mu\nu\lambda\rho}) 
\gamma^\lambda L \sum_f
\int\frac{d^4p}{(2\pi)^4} {\rm Tr}\; \left[ 
\gamma^\rho (X_f+Y_f\gamma_5) iS_F^{(f)}(p) \right] 
\end{eqnarray}
for the contribution to the vertex function.
For the background-dependent part, this yields
	\begin{eqnarray}\label{Gam2B}
\Gamma'{}^{(2B)}_{\mu\nu} 
= - \,\frac{g_Z^2}{M_Z^2}a'_{\mu\nu\lambda\rho} 
\gamma^\lambda L \sum_f
\int\frac{d^4p}{(2\pi)^3} {\rm Tr}\; \left[ 
\gamma^\rho (X_f+Y_f\gamma_5) (\rlap/p+m_f) \right] \delta(p^2-m_f^2)
\eta_f(p) 
	\end{eqnarray}
which, by taking the trace 
and using Eqs.\ (\ref{bZ}) and (\ref{intp}), can be reduced to
\begin{eqnarray}\label{Gam2Bfinal}
\Gamma'{}^{(2B)}_{\mu\nu} & = &
- b^{(Z)}a'_{\mu\nu\lambda\rho} 
\gamma^\lambda L v^\rho \nonumber\\
& = & - b^{(Z)} 
[\eta_{\mu\nu} \rlap/v + \gamma_\mu v_\nu + \gamma_\nu v_\mu ]L \,.
\end{eqnarray}

\subsubsection{Diagrams (C) and (D)}
{}For these diagrams we use the vertices given in Eqs.\ (\ref{LZint})
and (\ref{Lnch}). The $O(1/M_Z^2)$ term is given by
	\begin{eqnarray}\label{Gam2CD}
-i\kappa\Gamma^{(2C)}_{\mu\nu} =
-  (-ig_Z)
\left(\frac{i}{M_Z^2}\right)(-i\kappa g_Z a_{\mu\nu\lambda\rho})
\gamma^\lambda L \sum_f
\int\frac{d^4p}{(2\pi)^4} {\rm Tr}\; \left[
\gamma^\rho (X_f+Y_f\gamma_5) iS_F^{(f)}(p) \right] \,,
	\end{eqnarray}
and, since $a_{\mu\nu\lambda\rho}$ is symmetric in the indices $\lambda$
and $\rho$, it follows that $\Gamma^{(2D)}_{\mu\nu} =  
\Gamma^{(2C)}_{\mu\nu}$.
Following the same steps that led to Eq.\ (\ref{Gam2Bfinal})
we find for the background dependent part
\begin{equation}\label{Gam2CDfinal}
\Gamma'{}^{(2C)}_{\mu\nu} = \Gamma'{}^{(2D)}_{\mu\nu} =
b^{(Z)}[\eta_{\mu\nu} \rlap/v + \frac{1}{2}(\gamma_\mu v_\nu + 
\gamma_\nu v_\mu) ]L \,.
\end{equation}

\subsection{$W$-exchange diagrams}
\subsubsection{Diagram (A)}
	%
We now consider the diagrams shown in
{}Fig.~\ref{fig:wdiagrams}. We begin with  diagram A, 
which corresponds to the expression
\begin{eqnarray}\label{GammaA}
-i\kappa \Gamma^{(1A)}_{\mu\nu} = 
\left(-\frac{ig}{\sqrt{2}}\right)^2 (-i\kappa)\left(\frac{i}{M_W^2}\right) 
\int\frac{d^4p}{(2\pi)^4} \gamma_\lambda L iS_F^{(e)}(p') 
V_{\mu\nu}(p,p')
iS_F^{(e)}(p)\gamma^\lambda L  \,,
\end{eqnarray}
where we have neglected the momentum dependence of the $W$ propagator.
Using the identity
\begin{eqnarray}\label{identity}
(\gamma^\lambda L)M(\gamma_\lambda L) = 
-(\gamma^\lambda L)\mbox{Tr}(M\gamma_\lambda L) \,,
\end{eqnarray}
which is valid for any $4\times 4$ matrix $M$, 
Eq.\ (\ref{GammaA}) can be written as
\begin{eqnarray}\label{Gam1A}
\Gamma^{(1A)}_{\mu\nu} = 
\left(\frac{ig^2}{2M_W^2}\right) \, \gamma^\lambda L 
\int\frac{d^4p}{(2\pi)^4} {\rm Tr} \left[ iS_F^{(e)}(p') 
V_{\mu\nu}(p,p')
iS_F^{(e)}(p)\gamma_\lambda L \right] \,.
\end{eqnarray}
This expression coincides with what is obtained from
Eq.\ (\ref{Gamma2a}) by
discarding the nucleon terms and then making there the substitutions
\begin{eqnarray}\label{ZtoW}
{g_Z^2\over M_Z^2} \to {g^2\over 2M_W^2} \,,\quad X_e \to {1\over 2}
\,, \quad Y_e \to - {1\over 2} \,.
\end{eqnarray}
With this observation, the result for the background-induced part of
$\Gamma^{(1A)}_{\mu\nu}$ can be immediately deduced to be
\begin{eqnarray}\label{Gam1Afinal}
\Gamma'{}^{(1A)}_{\mu\nu} = \Lambda^{(W)}_{\mu\nu} -
b_e \rlap/{v} L\eta_{\mu\nu} \,,
\end{eqnarray}
where
\begin{eqnarray}\label{LambdaWfinal}
\Lambda^{(W)}_{\mu\nu} = 
\frac{g^2}{4M_W^2}(\gamma^\lambda L)
\int\frac{d^3P}{2E_e(2\pi)^3}
\left\{(f_e - f_{\overline e})\left[
\frac{N^{(1)}_{\mu\nu\lambda}(p_e,q)}{q^2 - 2p_e\cdot q} + 
(q\rightarrow -q)\right]\right.\nonumber\\
+ 
\left. (f_e + f_{\overline e})\left[
\frac{N^{(2)}_{\mu\nu\lambda}(p_e,q)}{q^2 - 2p_e\cdot q} -
(q\rightarrow -q)\right]
\right\} \,.
\end{eqnarray}
%

\subsubsection{Other diagrams}
The steps to follow in the calculation of the other diagrams
of Fig.~\ref{fig:wdiagrams} are very similar
to those for the corresponding diagrams of Fig.\ \ref{fig:zdiagrams}.
Omitting the details here, the background-dependent terms
are given by the formulas
\begin{eqnarray}\label{Gam1BCDprime}
-i\kappa\Gamma^\prime{}^{(1B)}_{\mu\nu} & = &
-\left(-\frac{ig}{\sqrt{2}}\right)^2
\left(\frac{i}{M^2_W}\right)^2
(i\kappa M^2_W a^\prime_{\mu\nu\alpha\beta})
\int\frac{d^4p}{(2\pi)^3}\delta(p^2 - m^2_e)\eta_e(p)
\gamma^\alpha L(\rlap{/}{p} + m_e)
\gamma^\beta L \nonumber \\
-i\kappa\Gamma^\prime{}^{(C)}_{\mu\nu} & = & 
-\left(-\frac{ig}{\sqrt{2}}\right)^2
\left(\frac{i}{M^2_W}\right)
(-i\kappa a_{\mu\nu\alpha\beta})
\int\frac{d^4p}{(2\pi)^3}\delta(p^2 - m_e^2)\eta_e(p)
\gamma^\alpha L(\rlap{/}{p} + 
m_e)\gamma^\beta L \,,
\end{eqnarray}
with $\Gamma^\prime{}^{(D)}_{\mu\nu} = \Gamma^\prime{}^{(C)}_{\mu\nu}$.
Using Eq.\ (\ref{intp}), after some straightforward algebra these reduce to
\begin{eqnarray}\label{Gam1BCD}
\Gamma'{}^{(1B)}_{\mu\nu} & = & b_e \left[-\eta_{\mu\nu} \rlap/v 
+ {1\over 2} (\gamma_\mu \rlap/v \gamma_\nu + 
\gamma_\nu \rlap/v \gamma_\mu) \right]L \nonumber\\
\Gamma'{}^{(1C)}_{\mu\nu} = 
\Gamma'{}^{(1D)}_{\mu\nu} & = &
\frac{1}{2}b_e \left[ 2\eta_{\mu\nu} \rlap/v - 
\frac{1}{2} (\gamma_\mu \rlap/v
\gamma_\nu + \gamma_\nu \rlap/v \gamma_\mu)\right]L \,.
\end{eqnarray}

\subsection{Complete vertex}
We have completed the calculation of the diagrams that
give rise to the background-induced part of the proper
vertex of neutrinos with the graviton.
The complete vertex function defined in Eq.\ (\ref{defGameff})
is obtained by adding these individual contributions.
For the diagrams of Fig.~\ref{fig:wdiagrams} the results
are given in Eqs.\
(\ref{Gam1Afinal}) and (\ref{Gam1BCD}), and adding them we
obtain
	\begin{eqnarray}\label{wexchange}
\Gamma'{}^{(1A)}_{\mu\nu} + \Gamma'{}^{(1B)}_{\mu\nu} +
\Gamma'{}^{(1C)}_{\mu\nu} + \Gamma'{}^{(1D)}_{\mu\nu}  = 
\Lambda^{(W)}_{\mu\nu}\,.
	\end{eqnarray}
For the diagrams of Fig.~\ref{fig:zdiagrams},
the sum of the individual results given in
Eqs.\ (\ref{Gam2Afinal}), (\ref{Gam2Bfinal}) and (\ref{Gam2CDfinal})
yields
\begin{eqnarray}\label{zexchange}
\Gamma'{}^{(2A)}_{\mu\nu} + \Gamma'{}^{(2B)}_{\mu\nu} +
\Gamma'{}^{(2C)}_{\mu\nu} + \Gamma'{}^{(2D)}_{\mu\nu} =
\Lambda^{(Z)}_{\mu\nu} \,. 
\end{eqnarray}
As we have already mentioned, $\Lambda^{(W)}_{\mu\nu}$ 
contributes to the vertex
only when the neutrino is $\nu_e$,
while $\Lambda^{(W)}_{\mu\nu}$ contributes equally 
for all the weak-doublet neutrinos, including the $\nu_e$.
In addition, the vertex function has the tree-level term
which, for chiral, left-handed neutrinos is given in Eq.\ (\ref{Vmunu}).

In this way we find that the complete effective vertex function for the
various neutrino flavors is given by
\begin{eqnarray}\label{Gammaeff}
\Gamma^{(\nu)}_{\mu\nu} & = & V^{(\nu)}_{\mu\nu} +
\Lambda_{\mu\nu} \,,
\end{eqnarray}
where
\begin{eqnarray}\label{Lambda}
\Lambda_{\mu\nu} = \left\{ \begin{array}{ll}
\Lambda^{(W)}_{\mu\nu} + \Lambda^{(Z)}_{\mu\nu} \quad & \mbox{for
$\nu_e$\,,} \\ 
\Lambda^{(Z)}_{\mu\nu} & \mbox{for $\nu_\mu,\nu_\tau$\,.} \end{array}
\right. 
\end{eqnarray}
Notice that the vertex function is symmetric in the indices $\mu,\nu$, as it
should be.

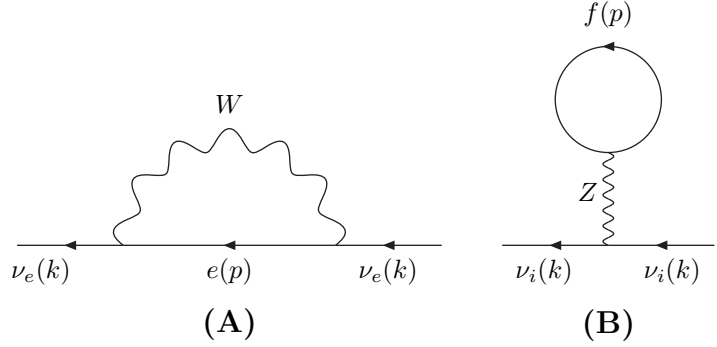
\begin{figure}
\begin{center}
%
%
\begin{picture}(180,130)(-90,-30)
\Text(0,-30)[c]{\large\bf (A)}
\ArrowLine(80,0)(40,0)
\Text(60,-10)[c]{$\nu_e(k)$}
\ArrowLine(40,0)(-40,0)
\Text(0,-10)[c]{$e(p)$}
\ArrowLine(-40,0)(-80,0)
\Text(-60,-10)[cr]{$\nu_e(k)$}
\PhotonArc(0,0)(40,0,180){4}{6.5}
\Text(0,50)[cb]{$W$}
\end{picture}
%
%
\begin{picture}(100,100)(-50,-30)
\Text(0,-30)[c]{\large\bf (B)}
\ArrowLine(40,0)(0,0)
\Text(35,-10)[cr]{$\nu_i(k)$}
\ArrowLine(0,0)(-40,0)
\Text(-35,-10)[cl]{$\nu_i(k)$}
\Photon(0,0)(0,35){2}{6}
\Text(-4,20)[r]{$Z$}
\ArrowArc(0,55)(20,-90,270)
\Text(0,85)[b]{$f(p)$}
\end{picture}
\caption[]{\sf One-loop diagrams for the self-energy of
neutrinos in a medium. Diagram (A) contributes only to
$\nu_e$, while diagram (B) contributes equally to $\nu_e$, $\nu_\mu$
and $\nu_\tau$.
\label{fig:selfenergy}}

\end{center}
\end{figure}

\subsection{Transversality of the vertex}
\label{subsec:transv}
Before proceeding to the explicit calculations of the neutrino vertex
using the one-loop formula given above, it is useful to check 
that the complete effective vertex satisfies 
the transversality condition
\begin{eqnarray}\label{transversality}
q^\mu\,\overline u_L(k')\, \Gamma^{(\nu)}_{\mu\nu}(k,k')\, u_L(k) = 
q^\nu\,\overline u_L(k')\, \Gamma^{(\nu)}_{\mu\nu}(k,k')\, u_L(k) = 0 \,.
\end{eqnarray}
Here, 
\begin{eqnarray}\label{konshell}
k^\mu & = & (\omega_K,\vec K)\nonumber\\
k^{\prime\mu} & = & (\omega_{K'},\vec K') \,,
\end{eqnarray}
where $\omega_K$ is the correct dispersion relation for a neutrino
mode propagating with momentum $\vec K$ in the medium,
and $u_L$ is the corresponding spinor.
As shown several years ago \cite{nr,pp,jfn}, 
the spinor $u_L(k)$ and the corresponding dispersion
relation are found by solving the effective Dirac equation
\begin{eqnarray}\label{eqmotion}
(\rlap/k - \Sigma_{\rm eff})\,u_L(k) = 0 \,,
\end{eqnarray}
where $\Sigma_{\rm eff}$ is the neutrino self-energy in the medium.
The chirality of the
neutrino interactions dictate that $\Sigma_{\rm eff}$ has the form
\begin{eqnarray}\label{Sigmaeff}
\Sigma_{\rm eff} = (a\rlap/k + b\rlap/v)L \,,
\end{eqnarray}
where, in general, $a$ and $b$ are functions of $\omega_K$ and $K$.
In this form, the dispersion relations implied by Eq.\ (\ref{eqmotion})
are given by
\begin{eqnarray}\label{disp}
\omega_K & = & K + {b(\omega_K,K)\over 1-a(\omega_K,K)} \nonumber\\
&&\nonumber\\
\overline\omega_K & = & K - \frac{b(-\overline\omega_K,K)}
{1-a(-\overline\omega_K,K)} \,,
\end{eqnarray}
with $\omega_K$ corresponding to the neutrino and $\overline\omega_K$
to the antineutrino,
which in the general case must be considered as implicit
equations that must be solved for $\omega_K$ and
$\overline\omega_K$ as functions of $K$.
In the context of our perturbative approach, the solutions
to Eq.\ (\ref{disp}) are given approximately by
\begin{eqnarray}\label{disp2}
\omega_K & = & K + [1 + a(K,K)]b(K,K)\nonumber\\
\overline\omega_K & = & K - [1 + a(-K,K)]b(-K,K)\,.
\end{eqnarray}
At the one-loop level, the neutrino self-energy in the presence of matter
is determined by calculating the diagrams shown in Fig.~\ref{fig:selfenergy},
which has been carried out in detail in the references cited.
The result is that, to order $1/M_W^2$, the parameter
$a$ vanishes while
\begin{eqnarray}\label{bnuall}
b_{\rm mat} = \left\{
\begin{array}{ll}
b_e + b^{(Z)} & \mbox{for $\nu_e$} \\
b^{(Z)} & \mbox{for $\nu_\mu,\nu_\tau$}
\end{array}\right. 
\end{eqnarray}
where $b_e$ and $b^{(Z)}$ are given in Eqs.\ (\ref{b}) and (\ref{bZ}),
respectively. In this case $\omega_K = K + b_{\rm mat}$
for the neutrinos, and it follows that the spinors satisfy the equation
\begin{eqnarray}\label{eqmotion2}
(\rlap/k - b_{\rm mat} \rlap/v)Lu_L(k) = 0 
\end{eqnarray}
and the relation
\begin{eqnarray}\label{k2formula}
k^2 = 2b_{\rm mat} k\cdot v - b^2 
\end{eqnarray}
holds.
These, together with the analogous relations for $u(k')$ and $k^{\prime 2}$, 
imply the useful formulas
\begin{eqnarray}\label{Ukrelations}
\overline u_L(k') \, \rlap/q \, u_L(k) & = & 0 \nonumber\\
k^2 - k^{\prime 2} & = & 2b_{\rm mat} q\cdot v \,.
\end{eqnarray}

We are now in the position to consider the transversality
property of the effective vertex. {}From Eq.\ (\ref{N}) it follows that
\begin{eqnarray}\label{qN}
q^\mu N^{(1)}_{\mu\nu\alpha} & = &
(2p\cdot q - q^2)[4p_\nu p_\alpha - 2q_\nu p_\alpha - p_\nu q_\alpha +
p\cdot q\eta_{\nu\alpha}] \nonumber \\
q^\mu N^{(2)}_{\mu\nu\alpha} & = &
(2p\cdot q - q^2)i\epsilon_{\nu\alpha\lambda\rho}q^\lambda p^\rho \,.
\end{eqnarray}
Then from Eq.\ (\ref{LambdaWfinal})
\begin{eqnarray}\label{qLambdaW}
q^\mu\Lambda^{(W)}_{\mu\nu} & = &
\frac{g^2}{2M_W^2}(\gamma^\alpha L)\int\frac{d^4p}{(2\pi)^3}
\delta(p^2 - m_e^2)\eta_e(p)
[2q_\nu p_\alpha  + p_\nu q_\alpha - p\cdot q\eta_{\nu\alpha}]\nonumber\\
& = & \frac{1}{2}b_e[2q_\nu\rlap/v + v_\nu\rlap/q - 
v\cdot q\gamma_\nu]L \,, 
\end{eqnarray}
and similarly
\begin{eqnarray}\label{qLambdaZ}
q^\mu\Lambda^{(Z)}_{\mu\nu} =
\frac{1}{2}b^{(Z)}[2q_\nu\rlap/v + v_\nu\rlap/q -
v\cdot q\gamma_\nu]L \,.
\end{eqnarray}
On the other hand, from Eq.\ (\ref{Vneutrino}), we obtain
\begin{eqnarray}\label{qGamma0}
q^\mu V^{(\nu)}_{\mu\nu} = 
\frac{1}{4}[\rlap/q (k + k')_\nu +
(k^2 - k^{\prime 2})\gamma_\nu]L - \frac{1}{2}q_\nu
[\rlap/k + \rlap/k\,']L \,.
\end{eqnarray}
{}From these, and using the relations  given
in Eqs.\ (\ref{eqmotion2}) and (\ref{Ukrelations}), it is easy to verify
that the transversality condition stated in Eq.\ (\ref{transversality})
is indeed satisfied.

%
%
\section{Neutrino index of refraction in a gravitational field}
\setcounter{equation}{0}
\label{sec:refindex}
\subsection{The self-energy of neutrinos in a static gravitational
field} 
Our aim in this section is to determine the correction
to the neutrino index of refraction in the presence of
a static gravitational field. To this end, 
let us consider the scattering of a neutrino
by a static gravitational potential
$\phi^{\rm ext}(\vec x)$ which is produced 
by a static mass density $\rho^{\rm ext}(\vec x)$.
Defining the Fourier transform of $\phi^{\rm ext}$ by
\begin{eqnarray}\label{phiq}
\phi^{\rm ext}(\vec x) = \int\frac{d^3q'} {(2\pi)^3} 
\phi (\vec q^{\,\prime}) e^{i\vec q^{\,\prime}\cdot\vec x} \,, 
\end{eqnarray}
with a similar definition for $\rho(\vec q^{\,\prime})$,
the corresponding metric is such that, in momentum space,
\begin{eqnarray}\label{hphirel}
h^{\mu\nu}(\vec q^{\,\prime}) = \frac{1}{\kappa} \phi(\vec
q^{\,\prime}) 
\left(2v^\mu v^\nu - \eta^{\mu\nu}\right) \,,
\end{eqnarray}
where we have used $-2\vec q^{\,2}\phi = \kappa^2\rho$.
The formula in Eq.\ (\ref{hphirel}) is the solution to the 
linearized field equation for the metric
with the static energy momentum tensor
$T^{\mu\nu} = v^\mu v^\nu \rho^{\rm ext}$.
Under the influence of such an external potential,
the off-shell $\nu$-$\nu$ transition amplitude is then
\begin{eqnarray}\label{Snunu}
S_{\nu\nu} = -i \kappa (2\pi)\delta(k^0 - k^{\prime 0})
(V^{(\nu)}_{\mu\nu}(k,k') +
\Lambda_{\mu\nu}(0,\vec{\cal Q}))h^{\mu\nu}(\vec k' - \vec k) \,.
\end{eqnarray}
In Eq.\ (\ref{Snunu}) we have indicated explicitly the fact
that the background part $\Lambda_{\mu\nu}(\Omega,\vec{\cal Q})$ 
of the vertex function does not depend on $k$ and 
$k^\prime$ separately, but only
on the variables $\Omega$ and $\vec{\cal Q}$ that are defined
by writing
\begin{eqnarray}\label{OmegaQ}
q^\mu = (k - k')^\mu = (\Omega,\vec{\cal Q})
\end{eqnarray}
in the rest frame of the medium.  Moreover,  we have
set $\Omega = 0$ as implied by the delta function 
in the right-hand side.
In addition, we now restrict ourselves to a uniform
field in space, which means that can write
\begin{eqnarray}\label{phiconst}
\phi(\vec k' - \vec k) = (2\pi)^3\delta^{(3)}(\vec k' - \vec k)
\phi^{\rm ext} \,.
\end{eqnarray}
The justification for this is the usual
one. Namely, we assume that we are working in a region
of space that is microscopically large but macroscopically small,
and therefore the external field is approximately constant
over it.  The macroscopic dependence of the external field
on $\vec x$ can then be restored at the end.

Thus, substituting Eq.\ (\ref{phiconst}) into Eq.\ (\ref{Snunu}), we obtain
\begin{eqnarray}\label{Smunuuniform}
S_{\nu\nu} = - i(2\pi)^4 \delta^{(4)}(k - k')
(V^{(\nu)}_{\mu\nu}(k,k) +
\Lambda_{\mu\nu}(0,\vec{\cal Q}\rightarrow 0))
\left(2v^\mu v^\nu - \eta^{\mu\nu}\right) \phi^{\rm ext} \,.
\end{eqnarray}
Identifying the gravitational contribution to
the self-energy by writing
\begin{eqnarray}\label{identifysigma}
S_{\nu\nu} = -i(2\pi)^4\delta^{(4)}(k - k')\Sigma_G(k) \,,
\end{eqnarray}
we then find
\begin{equation}\label{SigmaG}
\Sigma_G = \Sigma_g + \Sigma^\prime_G \,,
\end{equation}
where
\begin{eqnarray}\label{sigmagrav}
\Sigma_g(k) & = & \phi^{\rm ext}
V^{(\nu)}_{\mu\nu}(k,k)
\left(2v^\mu v^\nu - \eta^{\mu\nu}\right) \nonumber\\
&&\nonumber\\
\Sigma^\prime_G(k) & = & \phi^{\rm ext}
\Lambda_{\mu\nu}(0,\vec{\cal Q}\rightarrow 0)
\left(2v^\mu v^\nu - \eta^{\mu\nu}\right) \,.
\end{eqnarray}
As indicated in Eq.\ (\ref{sigmagrav}), the gravitational
contribution to the self-energy has two
parts. 
We consider first $\Sigma_g(k)$, which is the contribution from the
gravitational field  when there is no background medium
present. Using the expression for $V_{\mu\nu}$ given in 
Eq.\ (\ref{Vneutrino}), we obtain
	\begin{eqnarray}
\Sigma_g(k) = \phi^{\rm ext} \left( \rlap/k + 2k\cdot v \rlap/v
\right) L \,,
	\end{eqnarray}
which in terms of the parametrization of the self-energy given in 
Eq.\ (\ref{Sigmaeff}) amounts to
	\begin{eqnarray}\label{agbg}
a_g = \phi^{\rm ext}\,, \qquad b_g = \phi^{\rm ext} 2k\cdot v \,.
	\end{eqnarray}

\subsection{The matter-induced gravitational contribution to the self
energy} 
Since the expression for $\Lambda_{\mu\nu}$ contains the factor 
$\gamma^\lambda L$, it is useful to define the vector
$t_\lambda$ by writing 
\begin{eqnarray}\label{t}
\left(2v^\mu v^\nu - \eta^{\mu\nu}\right)
\Lambda_{\mu\nu}(\Omega,\vec{\cal Q}) \equiv
(\gamma^\lambda L) t_\lambda(\Omega,\vec{\cal Q}) \,,
\end{eqnarray}
in terms of which
\begin{equation}\label{sigmagravnu}
\Sigma^\prime_G(k) = \phi^{\rm ext}\gamma^\lambda
t_\lambda(0,\vec{\cal Q}\rightarrow 0) \,.
\end{equation}
In an isotropic background, which we have assumed by writing
the distribution functions as in Eq.\ (\ref{fe}), $t_\alpha$
can be expressed in the form
\begin{equation}\label{tform}
t_\lambda = A_Gq_\lambda + B_G \tilde v_\lambda \,,
\end{equation}
where we have defined
\begin{equation}\label{utilde}
\tilde v^\lambda\equiv v^\lambda - \frac{v\cdot q}{q^2}q^\lambda \,.
\end{equation}
Since $q\cdot\tilde v = 0$, the scalar functions can be calculated
by using
\begin{eqnarray}\label{AGBG}
A_G & = & \frac{1}{q^2}q\cdot t \,,\nonumber\\
B_G & = & \frac{1}{\tilde v^2}\tilde v\cdot t \,.
\end{eqnarray}
If $A_G(0,\vec{\cal Q})$ is not singular in the limit
$\vec{\cal Q}\rightarrow 0$, 
then Eqs.\ (\ref{sigmagravnu}) and (\ref{tform}) imply that
\begin{equation}\label{sigmagravnu2}
\Sigma^\prime_G =
\phi^{\rm ext}
B_G(0,\vec{\cal Q}\rightarrow 0)\rlap{/}{v}L \,,
\end{equation}
so that only $B_G$ needs to be evaluated.  To prove
that this is the case, notice from Eq.\ (\ref{N}) that
\begin{equation}\label{N2identity}
v^\lambda\left(2v^\mu v^\nu - \eta^{\mu\nu}\right)
N^{(2)}_{\mu\nu\lambda} =
q^\lambda\left(2v^\mu v^\nu - \eta^{\mu\nu}\right)
N^{(2)}_{\mu\nu\lambda} = 0\,.
\end{equation}
Therefore, $A_G$ and $B_G$ are given by
\begin{eqnarray}\label{AGBG2}
A_G & = & \sqrt{2}G_F\left\{ \begin{array}{ll} 
A_e + \sum_f X_f A_f \quad & \mbox{for $\nu_e$\,,} \\ \\ 
\sum_f X_f A_f & \mbox{for $\nu_\mu,\nu_\tau$\,,} 
\end{array} \right. \nonumber\\
B_G & = & \sqrt{2}G_F\left\{ \begin{array}{ll} 
B_e + \sum_f X_f B_f \quad & \mbox{for $\nu_e$\,,} \\ \\ 
\sum_f X_f B_f & \mbox{for $\nu_\mu,\nu_\tau$\,,} 
\end{array} \right. 
\end{eqnarray}
with
\begin{eqnarray}\label{AfBf}
A_f & = & \frac{1}{q^2}
\int\frac{d^3P}{(2\pi)^3 2E_f}
(f_f - f_{\overline f})\left[
\frac{I_a(\Omega,\vec{\cal Q})}{q^2 - 2p_f\cdot q} -
(q\rightarrow -q)\right] \,,\nonumber\\
B_f & = & \frac{1}{\tilde v^2}
\int\frac{d^3P}{(2\pi)^3 2E_f}
(f_f - f_{\overline f})\left[
\frac{I_b(\Omega,\vec{\cal Q})}{q^2 - 2p_f\cdot q} +
(q\rightarrow -q)\right] \,,
\end{eqnarray}
where we have defined
\begin{eqnarray}\label{Iab}
I_a & = & q^\lambda\left(2v^\mu v^\nu - \eta^{\mu\nu}\right)
N^{(1)}_{\mu\nu\lambda} \,,\nonumber\\
I_b & = & \tilde v^\lambda\left(2v^\mu v^\nu - \eta^{\mu\nu}\right)
N^{(1)}_{\mu\nu\lambda} \,.
\end{eqnarray}
{}From the expression given in Eq.\ (\ref{N}) for $N^{(1)}_{\mu\nu\lambda}$
it is straightforward to see that 
\begin{equation}\label{Iastatic}
I_a(0,\vec{\cal Q}) =
\left({\cal Q}^2 - 2\vec P\cdot\vec{\cal Q}\right)
\left(8E_f^2 - 4m_f^2 -2\vec P\cdot\vec{\cal Q}\right) \,.
\end{equation}
Using this in Eq.\ (\ref{AfBf}), it follows that 
$A_f(0,\vec{\cal Q})$ is proportional to the integral
of $\vec P\cdot\vec{\cal Q}$, which is zero for an isotropic distribution.
Therefore
\begin{equation}\label{Afstatic}
A_f(0,\vec{\cal Q}) = 0 \,,
\end{equation}
which proves Eq.\ (\ref{sigmagravnu2}).

To evaluate $B_f(0,\vec{\cal Q}\rightarrow 0)$,
we use Eq.\ (\ref{Iab}) and the definition of $N^{(1)}_{\mu\nu\alpha}$
given in Eq.\ (\ref{N}) to obtain, after rearranging some terms,
\begin{equation}\label{Ibstatic}
I_b(0,\vec{\cal Q}) = 2E_f(8E_f^2 - 4m_f^2 - 2{\cal Q}^2) -
6E_f(2\vec P\cdot\vec{\cal Q} - {\cal Q}^2)\,.
\end{equation}
Using this in Eq.\ (\ref{AfBf}),
\begin{eqnarray}
B_f (0,\vec{\cal Q}) \equiv -3(n_f - n_{\overline f})
+ \int\frac{d^3P}{(2\pi)^3}
\left[
\frac{F - 2{\cal Q}^2(f_f - f_{\overline f})}
{-{\cal Q}^2 + 2\vec P\cdot \vec{\cal Q}} + 
(\vec{\cal Q} \rightarrow -\vec{\cal Q}) \right] \,,
\end{eqnarray}
with
\begin{eqnarray}\label{F}
F\equiv 4(2E_f^2 - m_f^2)(f_f - f_{\bar f}) \,.
\end{eqnarray}
To evaluate this integral, we make the change of variables
$\vec P\rightarrow \vec P + \frac{1}{2}\vec{\cal Q}$ in the first
term, and $\vec P\rightarrow \vec P - \frac{1}{2}\vec{\cal Q}$
in the second one, remembering that the distribution functions
and $F$ are functions of $\vec P$.  
This procedure yields
\begin{equation}\label{Bf0}
B_f (0,\vec{\cal Q}\rightarrow 0) = 
J_f + O\left({\cal Q}^2\right) \,,
\end{equation}
where
\begin{eqnarray}\label{J0}
J_f & = & -3(n_f - n_{\overline f}) + 
\int\frac{d^3P}{(2\pi)^3}
\left(
\frac{\vec{\cal Q}\cdot\vec\nabla_P F}{2\vec P\cdot\vec{\cal Q}}
\right) \,,\nonumber\\
& = & -3(n_f - n_{\overline f}) +
\int\frac{d^3P}{(2\pi)^3 2E_f} \; \frac{dF}{dE_f} \,.
\end{eqnarray}
In arriving at Eq.\ (\ref{J0}) we have used the definition in Eq.\ (\ref{ne}),
and in writing the second equality we have used 
the fact the the function $F$ depend on $\vec P$
only through $E_f$.  {}From Eqs.\ (\ref{AGBG2}), (\ref{sigmagravnu2}) 
and (\ref{Bf0}) we finally  obtain
\begin{equation}\label{SigmaGfinal}
\Sigma^\prime_G(k) = b_G\rlap{/}{v}
\end{equation}
where
\begin{eqnarray}\label{bGfinal}
b_G & = & \phi^{\rm ext} \sqrt{2}G_F \times
\left\{ \begin{array}{ll}
J_e +
\sum_f X_f J_f  & \mbox{for $\nu_e$}\,,\\
&\\
\sum_f X_f J_f   & \mbox{for
$\nu_\mu,\nu_\tau$} 
\end{array}\right.  
\end{eqnarray}

\subsection{Dispersion relations}
\label{subsec:disprelations}
We are now in the position
to determine the dispersion relation for the neutrinos
in the medium in the presence of a static gravitational field.
Referring back to Eq.\ (\ref{Sigmaeff}), the results we have obtained
can be summarized by writing
\begin{eqnarray}\label{abtotal}
a & = & a_g \nonumber\\
b &=& b_{\rm mat} + b_g + b_G \,.
\end{eqnarray}
The terms $a_g$ and $b_g$, which are given in Eq.\ (\ref{agbg}),
represent the contribution from pure gravity and
are present even in the case that there is no matter.
On the other hand, the term $b_{\rm mat}$,
given in Eq.\ (\ref{bnuall}), arises due to the
presence of the background  medium, independently of
whether a gravitational field is present or not. 
Finally, $b_G$, given in Eq.\ (\ref{bGfinal}),
is the contribution that arises due to the simultaneous  
presence of matter and the external gravitational field.
Thus, substituting Eq.\ (\ref{abtotal}) in Eq.\ (\ref{disp2}),
and keeping in mind that we are allowed to retain only terms
that are linear in the external gravitational potential,
we obtain
\begin{equation}\label{gravdisp}
\omega_K = K + 2K\phi^{\rm ext} + 
b_{\rm mat} + \left(\phi^{\rm ext}b_{\rm mat} + b_G \right)\,.
\end{equation}
The corresponding formula for $\overline\omega_K$ is obtained
from this by reversing the sign in front
of $b_{\rm mat}$ and $b_G$.
Apart from the first term in the right-hand side Eq.\ (\ref{gravdisp}), 
which corresponds just to the vacuum dispersion relation, the other terms
have the following meaning.

The second term has a purely gravitational origin.
If we take it by itself and neglect the effects of the medium, 
it gives rise to the dispersion relation
\begin{equation}\label{drgrav}
\overline\omega_K = \omega_K = K(1 + 2\phi^{\rm ext})\,,
\end{equation}
which is equivalent to say that the neutrino and antineutrino
acquire an index of refraction given by
\begin{equation}\label{refindexgrav}
{\cal N}_0 \equiv 1 - 2\phi^{\rm ext}\,.
\end{equation}
This is the same result that is obtained by solving
the equation
\begin{equation}\label{eikonal}
k_\mu k_\nu g^{\mu\nu} = 0\,,
\end{equation}
which is the appropriate equation to solve in order to determine
the dispersion relation for the photon or a massless scalar
particle.
That purely gravitational term also gives rise
to effects analogous to the gravitational red shift and bending of light.
For example, consider a neutrino beam propagating along the $z$ direction
in the vicinity of a massive body of mass $M$ situated
at the origin, with an impact parameter $b$.
In the eikonal approximation, $\omega_K(t)$ and
$\vec K(\vec x)$ satisfy the Hamilton equation
\begin{equation}\label{hamilton}
\frac{\partial\vec K}{\partial t} = -\vec\nabla\omega_K \,,
\end{equation}
which can be used to obtain the transverse component $K_x$ that the momentum
develops as a consequence of the influence of the gravitational
field.  Substituting $\omega_K = (1 + 2\phi^{\rm ext})K$
in Eq.\ (\ref{hamilton}), we can write
\begin{equation}\label{Kx}
\frac{K_x}{K} = -2\int_{-\infty}^{\infty}dt\frac{\partial\phi^{\rm ext}(r)}
{\partial b} \,,
\end{equation}
where the position vector is $\vec r = (b,0,t)$ and 
$\phi^{\rm ext}(r) = -GM/r$.  From this we obtain
the bending formula for neutrinos
\begin{equation}\label{bending}
\frac{K_x}{K} = -\,\frac{4GM}{b}\,,
\end{equation}
which is the same result as the corresponding formula for photons.

In contrast to the purely gravitational terms, the matter-dependent terms
contribute with opposite signs to the neutrino and antineutrino
indices of refraction.  
The term $b_{\rm mat}$ on the right side of 
Eq.\ (\ref{gravdisp}), which denotes 
the matter contribution in the absence of a
gravitational field, is the usual Wolfenstein term.  
It is flavor dependent, and it is the
origin of the MSW mechanism for neutrino oscillations in matter. 
The remaining terms are the new contributions we set out to determine.
One of them is exactly the Wolfenstein term multiplied by the
external gravitational potential. Since we are considering weak
gravitational fields, that term is much smaller than the
usual Wolfenstein term and can be neglected.
Thus, the dispersion relation reduces to
\begin{equation}\label{gravdisp2}
\omega_K = K + 2\phi^{\rm ext}K + 
b_{\rm mat}  + b_G \,.
\end{equation}
which can be equivalently stated in terms of the index of refraction
\begin{eqnarray}\label{refindex}
{\cal N}_\nu & \equiv & \frac{K}{\omega_K}\nonumber\\
& = & {\cal N}_0 -
\frac{b_{\rm mat}}{K} -
\frac{b_G}{K}\,.
\end{eqnarray}
The term that we have denoted by $b_G$
is more interesting since it can have a non-trivial
dependence on the temperature and density of the background material. 
It is flavor dependent as well, and it arises because of the presence of
both matter and gravitational fields.  
In order to be able to consider its possible physical
effects, in the ensuing section we will estimate its
magnitude for various physical conditions of the background
medium.

\subsection{Estimates of the matter-gravitational effects}
\label{subsec:estimates}

The question we wish to address here is how large are the
matter-gravitational effects, represented by the term $b_G$
in Eq.\ (\ref{refindex}), relative to the Wolfenstein term $b_{\rm mat}$.
To answer this question we need an estimate of the integral $H_f$,
which we have carried out in the Appendix, and also make
some assumptions about the composition of the background for the
possible physical situations of interest.

As an example, let us consider a medium such as the Sun
or a supernova.  In these, the conditions are such that
the nucleons are non-relativistic.  Therefore, using the results given in
Eqs.\ (\ref{Jfclassnr}) and (\ref{Jfdeglimiting}), we can write for the nucleons
($N = p,n$)
\begin{equation}\label{JN}
J_N =
\left\{
\begin{array}{ll}
- \beta m_N n_N & \mbox{classical nucleon gas}\\
&\\
- \, \frac{\textstyle 3n_N}{\textstyle v_{FN}^2} & 
\mbox{degenerate nucleon gas\,,}
\end{array}\right. 
\end{equation}
where $v_{FN}$ stands for the Fermi velocity of the nucleon gas.

Regarding the electrons, if they are relativistic then
from Eq.\ (\ref{Jfur}) we have $J_e = -5 n_e$.  In this case,
the electron contribution in Eq.\ (\ref{bGfinal}) is of the
order of the Wolfenstein term multiplied by $\phi^{\rm ext}$,
and therefore it is unimportant. On the other hand,
this is not necessarily the case for
a non-relativistic electron gas.  In analogy with
Eq.\ (\ref{JN}) for this case
\begin{equation}\label{Jenr}
J_e =
\left\{
\begin{array}{ll}
- \beta m_e n_e \quad & \mbox{classical non-relativistic electron
gas\,,}\\ &\\
- \, \frac{\textstyle 3n_e}{\textstyle v_{Fe}^2} & \mbox{degenerate
non-relativistic 
electron gas\,.}
\end{array}\right.
\end{equation}

To assess further the possible importance of the matter-gravitational
contributions to the index of refraction, we also need some
knowledge of the magnitude of the gravitational potential
that could be involved.  If we take as a guiding value the potential
at the surface of the Sun, 
\begin{equation}\label{phisun}
\phi_\odot = -2 \times 10^{-6} \,,
\end{equation}
we find that the matter-gravitational contributions could
be relevant under the appropriate conditions.
This is particularly true for the nucleon terms, which 
have the enhancement factors $\beta m_N$ or
$v_{FN}^{-2}$ in the classical and the degenerate cases,
respectively.  This is not unexpected, since the gravitational
potential couples more strongly to the more massive particles.

As we have already mentioned, and as Eq.\ (\ref{bGfinal})
clearly indicates,
the nucleon contributions to $b_G$ are the same for all the neutrino
flavors, and therefore they are irrelevant for neutrino
oscillations involving only weak-doublet neutrinos.
On the other hand, they are relevant for oscillation phenomena in which
sterile neutrinos participate and, as we have shown above,
they may be important.  Furthermore, $b_{\rm mat}$ and $b_G$
have a different dependence on the neutrino coordinate as it 
propagates through the medium, a property that may have 
also distinctive implications.

%
%
\section{Conclusions}
\setcounter{equation}{0}
\label{sec:conclusions}
In this work we have determined the effects of a static
gravitational potential on the neutrino index of
refraction in matter.  This has been done by first carrying out 
the one-loop calculation of the matter-induced gravitational
couplings of the neutrinos, and then by determining the
corrections to the neutrino self-energy that such couplings imply
in the presence of the gravitational potential.

As a consistency check of the one-loop formulas, we showed
explicitly that the effective gravitational vertex of the
neutrino is transverse.  This required that the correct
dispersion relation and wavefunction associated with
the external neutrino modes be used, and not their vacuum
counterpart.  We emphasize again that it is crucial
that, in the calculation of the one-loop diagrams,
we have used the full off-shell formula for the
tree-level gravitational vertex function of the internal fermions
in the loop, and not the on-shell limit that is customarily quoted.

As indicated in Section\ \ref{sec:refindex}, the matter-gravitational
contribution to the neutrino index of refraction 
could be relevant in the context of matter-enhanced
neutrino oscillations, and in particular in phenomena 
involving the so-called sterile neutrinos.
This may occur not only because their magnitude could be non-negligible, but
also because they have a different dependence on the coordinate
compared with the standard Wolfenstein term.
Whether or not these gravitational effects can lead to interesting observable
consequences in specific  contexts, such as the supernova or the
Solar neutrino problem, is an open question that needs
further detailed study.

The results presented
here indicate that such studies could be well motivated,
and our work sets down the arena to carry them out 
on firm grounds and in a systematic fashion.

\paragraph*{Note added:}
After this paper was submitted for publication, 
the work by Piriz, Roy and Wudka \cite{piriz} was 
brought to our attention,
in which the tree-level gravitational coupling
of the neutrino is also considered without violating
the Equivalence Principle.  However, these authors assume
an intrinsic magnetic moment of the neutrino,
and they concern themselves with the effect of the 
(vacuum) gravitational interactions on the magnetic spin flip 
oscillations.  They do not consider the effect of matter
on the gravitational neutrino interactions, which is the focus
of the present work.

\begin{center}
ACKNOWLEDGEMENT
\end{center}
The work of J.F.N. has been partially supported by the
US National Science Foundation Grant PHY-9600924.

%
%
\appendix

\section{Evaluation of $J_f$}
\setcounter{equation}{0}
\label{appendix}
We consider the evaluation of the quantity $J_f$ defined
in Eq.\ (\ref{J0}) for various conditions of the background fermion gas.
That expression for $J_f$ can be rewritten
by taking the derivative of the function
$F$ given in Eq.\ (\ref{F}) and using Eq.\ (\ref{ne}).  In this way Eq.\ (\ref{J0}) becomes
\begin{equation}\label{J0final}
J_f = (n_f - n_{\bar f}) + H_f
\end{equation}
where 
\begin{equation}\label{Kf}
H_f = 4\int\frac{d^3P}{(2\pi)^3} \; \left( E_f -
{m_f^2 \over 2E_f} \right) {d\over dE_f} (f_f - f_{\bar f}) \,.
\end{equation}
In order to carry out this integral, we consider the
following illustrative cases.

\subsection{Classical non-relativistic gas}
In this case, we can put $f_{\overline f}\simeq 0$ and use
\begin{equation}\label{fclassnr}
\frac{df}{dE_f} \simeq -\beta f \,.
\end{equation}
Using these and remembering
that we can approximate $E_f \simeq m_f$ in the integrand,
we then obtain from Eqs.\ (\ref{J0final}) and (\ref{Kf})
\begin{equation}\label{Jfclassnr}
J_f \simeq H_f \simeq - \beta m_f n_f \,,
\end{equation}
where we have used the fact that $\beta m_f \gg 1$ for a non-relativistic gas.

\subsection{Degenerate gas at zero temperature}
For this case, the distribution function is given by
\begin{eqnarray}
f = \Theta(E_F-E) \,, \qquad \overline f \approx 0\,,
\end{eqnarray}
where $\Theta$ denotes the step function and $E_F$ is the Fermi
energy. Thus $df/dE = -\delta(E-E_F)$. Putting this back into
Eq.\ (\ref{Kf}) and using $P_F^3=3\pi^2 n_f$ for the Fermi
momentum,  we obtain
\begin{equation}\label{Kfdeg}
H_f = - \left[ 6n_f + m_f^2 \left( {3n_f \over \pi^4}
\right)^{1/3} \right] \,,
\end{equation}
which in turn implies
\begin{equation}\label{Jfdeg}
J_f = - \left[ 5n_f + 
m_f^2 \left( {3n_f \over \pi^4}\right)^{1/3} \right] \,.
\end{equation}
No assumption has been made here about whether or not the gas is
non-relativistic.  However, in the relativistic or non-relativistic
limits, this formula reduces to
\begin{equation}\label{Jfdeglimiting}
J_f =
\left\{
\begin{array}{ll}
-5n_f \quad & \mbox{relativistic gas} \\
& \\
- \, \frac{\textstyle 3n_f}{\textstyle v_F^2} & \mbox{non-relativistic
gas} 
\end{array}\right.
\end{equation}
where we have used $p_F = m_f v_F$ for the non-relativistic case.

\subsection{Ultra-relativistic gas}

In this case we neglect the mass of the background particles and therefore
we approximate Eq.\ (\ref{Kf}) by
\begin{equation}\label{Kfur}
H_f = 4\int\frac{d^3P}{(2\pi)^3}P{d\over dP} (f_f - f_{\bar f})\,.
\end{equation}
By carrying out a partial integration and using Eq.\ (\ref{ne}) 
this is equivalent to
\begin{equation}\label{Kfurfinal}
H_f = -6(n_f - n_{\overline f}) \,,
\end{equation}
and therefore
\begin{equation}\label{Jfur}
J_f = -5(n_f - n_{\overline f}) \,.
\end{equation}


\begin{thebibliography}{99}
\bibitem{footnote1} This notation and that for the vierbein fields
should not be the source of confusion, firstly because the vierbeins
have two indices while the velocity has only one, and secondly because
the vierbeins will not be used in the rest of the paper.
\end{thebibliography}
\end{document}